\documentclass[aps,prapplied,reprint,superscriptaddress]{revtex4-2}

\usepackage{mathtools}
\usepackage{siunitx}
\usepackage{comment}
\usepackage{subcaption} 
\usepackage{bm}
\usepackage{graphicx}

\begin{document}

\preprint{FERMILAB-PUB-26-0351-SQMS}
\preprint{DESY-26-xxx}
 
\title{\boldmath Cryogenic RF characterization of the MAGO cavity for high-frequency gravitational-wave detection}

% \author{Can \surname{Dokuyucu}$^a$}
\author{Can \surname{Dokuyucu}}
\affiliation{Deutsches Elektronen-Synchrotron DESY,\\
 Notkestraße 85, 22607 Hamburg, Germany}

% \author{Bianca \surname{Giaccone}$^a$}
\author{Bianca \surname{Giaccone}}
\email{giaccone@fnal.gov}
\affiliation{Fermi National Accelerator Laboratory,\\
Kirk and Pine St, Batavia, IL 60510, United States}

% \author{Tom \surname{Krokotsch}$^a$}
\author{Tom \surname{Krokotsch}}
\affiliation{Universität Hamburg,\\
Luruper Chaussee 149, 22761 Hamburg, Germany}

% \author{Giovanni \surname{Marconato}$^a$}
\author{Giovanni \surname{Marconato}}
\email{giovanni.marconato@desy.de}
\affiliation{Universität Hamburg,\\
Luruper Chaussee 149, 22761 Hamburg, Germany}

% \author{Krisztian \surname{Peters}$^a$}
\author{Krisztian \surname{Peters}}
\email{krisztian.peters@desy.de}
\affiliation{Deutsches Elektronen-Synchrotron DESY,\\
 Notkestraße 85, 22607 Hamburg, Germany}

% \author{Marc \surname{Wenskat}$^a$}
\author{Marc \surname{Wenskat}}
\email{marc.wenskat@desy.de}
\affiliation{Deutsches Elektronen-Synchrotron DESY,\\
 Notkestraße 85, 22607 Hamburg, Germany}
\affiliation{Universität Hamburg,\\
Luruper Chaussee 149, 22761 Hamburg, Germany}

\author{Julien Branlard}
\affiliation{Deutsches Elektronen-Synchrotron DESY,\\
 Notkestraße 85, 22607 Hamburg, Germany}
 
\author{Vijay Chouhan}
 \affiliation{Fermi National Accelerator Laboratory,\\
Kirk and Pine St, Batavia, IL 60510, United States}

\author{Ivan Gonin}
 \affiliation{Fermi National Accelerator Laboratory,\\
Kirk and Pine St, Batavia, IL 60510, United States}

\author{Anna Grassellino}
 \affiliation{Fermi National Accelerator Laboratory,\\
Kirk and Pine St, Batavia, IL 60510, United States}

\author{Wolfgang Hillert}
\affiliation{Universität Hamburg,\\
Luruper Chaussee 149, 22761 Hamburg, Germany}

\author{Timergali Khabiboulline}
 \affiliation{Fermi National Accelerator Laboratory,\\
Kirk and Pine St, Batavia, IL 60510, United States}
 
\author{Oleksandr Melnychuk}
 \affiliation{Fermi National Accelerator Laboratory,\\
Kirk and Pine St, Batavia, IL 60510, United States}

\author{Gudrid Moortgat-Pick}
\affiliation{Deutsches Elektronen-Synchrotron DESY,\\
 Notkestraße 85, 22607 Hamburg, Germany}
\affiliation{Universität Hamburg,\\
Luruper Chaussee 149, 22761 Hamburg, Germany}

\author{Alexandr Netepenko}
 \affiliation{Fermi National Accelerator Laboratory,\\
Kirk and Pine St, Batavia, IL 60510, United States}

\author{Sam Posen}
\affiliation{Fermi National Accelerator Laboratory,\\
Kirk and Pine St, Batavia, IL 60510, United States}

\date{\today}

\begin{abstract}
Superconducting radio-frequency (SRF) cavities are promising resonant sensors for gravitational-wave detection in the kHz-MHz frequency range. We report the cryogenic RF characterization of a prototype superconducting niobium cavity with an unconventional geometry designed for narrow electromagnetic mode separation. Following an adapted surface preparation procedure, cryogenic tests were performed at Fermilab and DESY at temperatures down to 2\,K.
Mechanical tuning at room temperature achieved a mode splitting of approximately 11\,kHz at cryogenic temperature. High electromagnetic quality factors consistent with previous prototype cavities were measured. The measurements further revealed phase transfer characteristics relevant for stable low-level RF control as well as indications of mode coupling potentially caused by one-point multipacting. In addition, first cryogenic measurements of the mechanical eigenmodes yielded mechanical quality factors significantly below commonly assumed theoretical values.
These results demonstrate the successful application of established SRF preparation and characterization techniques to a non-standard resonator geometry and provide important experimental input for the development of future SRF-based gravitational-wave detectors.
\end{abstract}

%\keywords{}

\maketitle
% \noindent $^a$ These authors contributed equally to this work.

%-------------------------------------------------------------------------------
%-------------------------------------------------------------------------------
\section{\label{sec:intro}Introduction}

%-------------------------------------------------------------------------------

The successful detection of gravitational waves has opened a new era in observational astronomy, with scientific attention now increasingly turning toward the exploration of novel frequency ranges beyond those accessible with current detectors. For the kHz–MHz range, a particularly promising avenue lies in the use of meter-scale instruments whose detectable frequencies can be tuned to match those of gravitational waves, thereby unlocking new detection possibilities that complement existing interferometric approaches. In this context, superconducting radio-frequency (SRF) cavities emerge as a compelling technology, offering a notable advantage over their optical counterparts: for instruments of comparable physical size, SRF cavities are capable of storing several orders of magnitude more electromagnetic energy than optical cavities, enhancing their potential as gravitational-wave detectors. 

In the detection concept explored in this work, an electromagnetic resonator is configured with two nearly degenerate modes in frequency, where RF power is injected into only one mode, the so called $0$-mode. An incoming gravitational wave with a frequency $\omega_g$ deforms the cavity walls, leading to a power transfer from the symmetric $0$-mode to the asymmetric signal or $\pi$-mode, which is maximized when the resonant condition $|\omega_\pi - \omega_0| = \omega_g$ is met. This process, in which a signal at the difference frequency between the two modes is used for readout, is commonly referred to as heterodyne detection.
 
In an experimental R\&D program this detection concept was explored within the MAGO proposal with the goal for a scaled-up experiment with 500 MHz cavities as a CERN-INFN collaboration \cite{Ballantini:2003nt, Ballantini:2004wd, Ballantini:2005am}. 
Although the final project was not funded, three SRF niobium cavities were built during the R\&D activities. The first cavity (a pill-box cavity) was used as a proof-of-principle experiment, which demonstrated the working principle and the development of an RF system to drive and read out the cavity with the necessary precision~\cite{Ballantini:2003nt, Ballantini:2004wd, Ballantini:2005am} for a mode separation in the MHz range. The second cavity was a spherical 2-cell cavity with two flanges on each cell and an optimized geometry, but with a fixed coupling and hence a fixed mode separation.
The third cavity was the identical spherical 2-cell cavity with two flanges on each cell (denoted {\tt PACO-2GHz-variable}) and an optimized geometry but with a mechanical tunable cell in between, with the goal of achieving mode separation in the 10 kHz range that is still adjustable - see Fig.\, \ref{fig:BCP+furnace}. 

In a recent effort, this R\&D program was revived with the {\tt PACO-2GHz-variable} cavity at the University of Hamburg, Deutsches Elektronen-Synchrotron (DESY) and at the Superconducting Quantum Materials and Systems (SQMS) Center at Fermi National Accelerator Laboratory (Fermilab). The initial work comprised a mechanical survey of the cavity and a characterization of its electromagnetic properties at room temperature, along with a plastic tuning procedure to achieve the target mode separation in the 10 kHz range \cite{Fischer:2025}. In this follow-up work, we turn to the RF properties of the cavity, as determined through several cryogenic test campaigns conducted at DESY and Fermilab, which provide key input for the expected operation and gravitational-wave sensitivity.
    
%-------------------------------------------------------------------------------
\section{\label{sec:treatment}Surface Preparation}

%-------------------------------------------------------------------------------
Given the little information available regarding the cavity fabrication and history, it was decided to inspect the inner volume to assess the surface conditions and plan for adequate surface treatment to achieve high quality factors. The cavity shape, thickness and fabrication materials, as described in \cite{Fischer:2025}, posed some limitations on the choice of surface removal method and heat treatment, as discussed later.

\subsection{Optical Inspection}
\begin{figure}[!htbp]
    \includegraphics[width=\columnwidth]{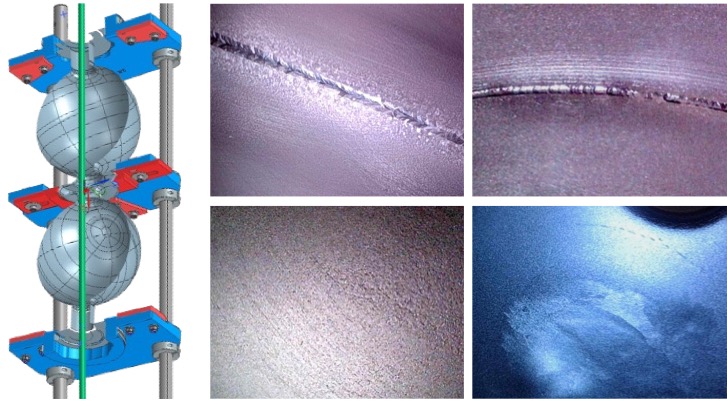}
    \caption{On the left, schematic of the cavity assembled for the boroscope inspection. The boroscope is supported via a rigid wand and slides through the cavity. The boroscope head is flexible and can be extended out of the wand to inspect the cavity surface. On the right, four images of the inner surface that show examples of the traces left from the fabrication process, such as scratches and rolling marks. In addition the bottom right image shows the inner side of one of the cavity cell dents (the `donut' indentation).}
    \label{fig:optical}
\end{figure}

Given the small aperture of the `beamtubes' ($<$\SI{4}{\centi\metre}), it was not possible to use the standard optical inspection tools developed for SRF cavities. As a result, the optical inspection was carried out with a lower resolution flexible boroscope inserted on the longitudinal axis of the cavity and also through the two cells transversal flanges. The purpose of the optical inspection was to investigate if the cavity had already received any surface treatment after fabrication, and if defects and damages were present on the inner surface.
From the morphology of the inner surface (shown in Fig.\,\ref{fig:optical}) it was clear that the cavity had undergone some degree of surface removal, however, it was not possible to determine if the surface removal had been carried out on the individual pieces prior to the welding, or on the cavity as a whole. Additionally, traces of the fabrication process (such as scratches) were still visible on the inner surface.

\subsection{Buffered Chemical Polishing (BCP)}
As a result of the optical inspection, it was decided that further surface removal was necessary to mitigate the impact of the fabrication scratches on the cavity performance. This includes both the desire of achieving a smooth inner surface to decrease risks of local field enhancement during the RF tests, and also, the goal of removing contaminants that may be left on the inner surface from the fabrication, and that may contaminate the cavity and furnace during the subsequent heat treatment.

\begin{figure}[!htbp]
    \includegraphics[width=\columnwidth]{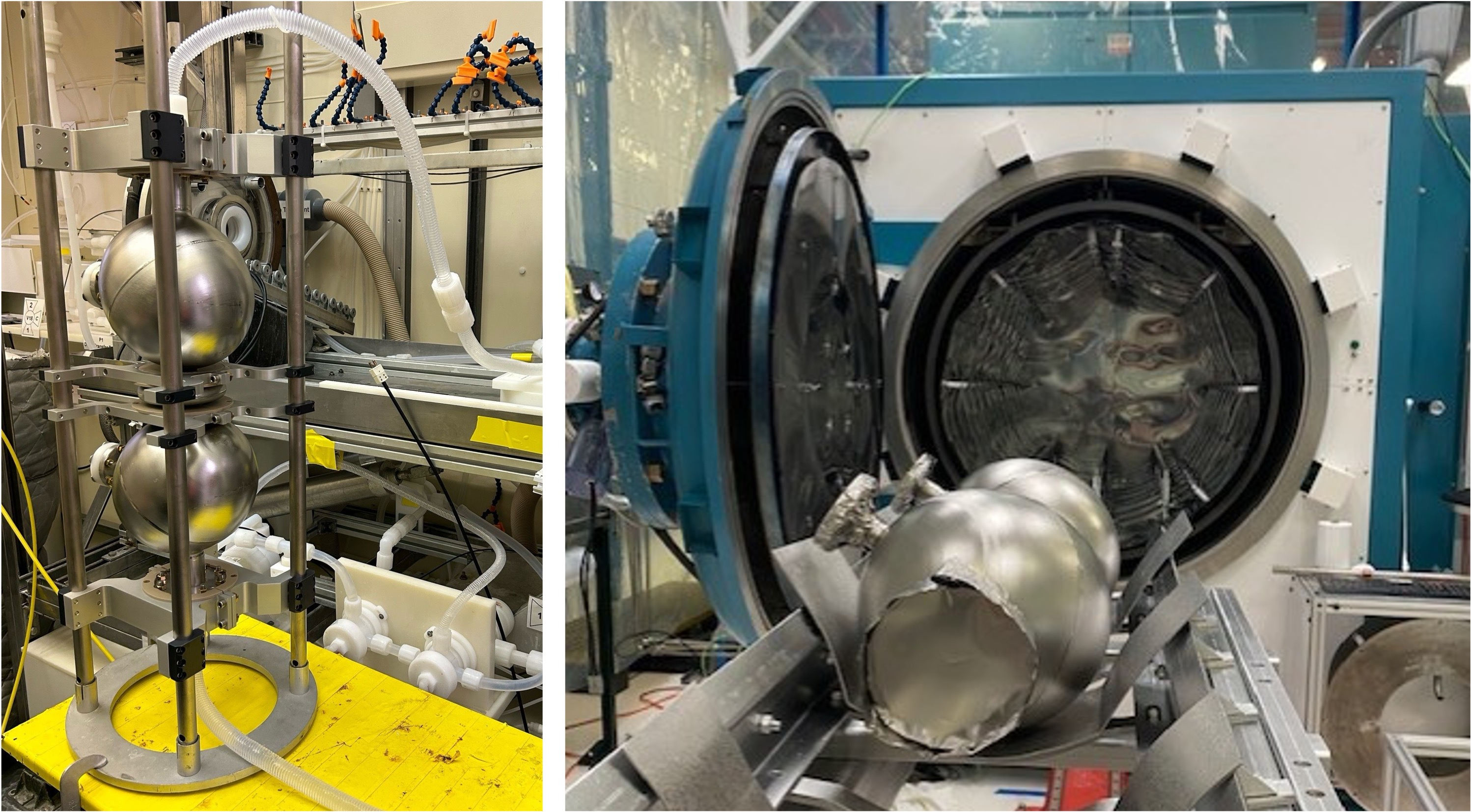}
    \caption{On the left, cavity installed in the frame and equipped with adaptors and flexible hoses for the BCP acid flow. On the right, cavity prepared for the 600\,$^{\circ}$C heat treatment. The cavity lays on niobium sheets bent in a U-shape and used as support structure. The cavity flanges are covered with niobium foil.}
    \label{fig:BCP+furnace}
\end{figure}

Due to the unique cavity shape with narrow `beamtube' apertures, narrow opening for the coupling cell, and wide almost-spherical cells, it was challenging to apply electropolishing (EP) to remove material from the inner surface, as EP would require the use of a cathode placed inside the cavity volume, as explained in \cite{chouhan:srf2023-tuptb042}. As an alternative, buffered chemical polishing (BCP) was employed. Given the shape of the cavity, we expected the material removal to be non-uniform and to be higher in areas such as the coupling cell, where the material thickness was already reduced to \SI{1}{\milli\metre}. Additionally, BCP is usually conducted with a rotational system that allows to rotate the cavity while the acid is flown through to avoid gas accumulation on the niobium surface, which can lead to surface defects and increased roughness. However, given the unique and non-straight cavity geometry, the rotational BCP was not applied. These constraints led to the decision to conduct only a flash BCP aiming for a removal of 4--\SI{6}{\micro\metre}. The BCP was conducted at the SRF facilities at Argonne National Laboratory, Fig.\,\ref{fig:BCP+furnace} on the left shows the setup used for the flash BCP.

\subsection{Heat Treatment}
The cavity flanges are made of stainless steel and are brazed on the niobium tubes. For this reason, the temperature of the heat treatment was limited to 600\,$^{\circ}$C to avoid contamination of the cavity and the furnace. The cavity was heated to 600\,$^{\circ}$C for 24 hours as degassing step, to remove the hydrogen that may have been introduced during the fabrication process and to relieve internal stresses introduced during the fabrication. %The heat treatment is done under vacuum and a Residual Gas Analyser is used to track the partial pressure of possible contaminants. During the {\tt PACO-2GHz-variable} heat treatment, no contaminant were measured, confirming that limiting the temperature to 600\,$^{\circ}$C successfully mitigates the risks of contamination. 
Fig.\,\ref{fig:BCP+furnace} on the right shows the cavity as prepared for the heat treatment.
% \begin{comment}
% \begin{figure}
%     \centering
%     \includegraphics[width=0.65\textwidth]{MAGO_Furnace.jpg}
%     \caption{{\tt PACO-2GHz-variable} cavity prepared for the 600\,$^{\circ}$C heat treatment. The cavity lays on niobium sheets bent in a U-shape and used as support structure. The cavity flanges are covered with niobium foil.}
%     \label{fig:furnace}
% \end{figure}
% \end{comment}
\subsection{High Pressure Rinsing (HPR)}
\begin{figure}[!htbp]
    \includegraphics[width=\columnwidth]{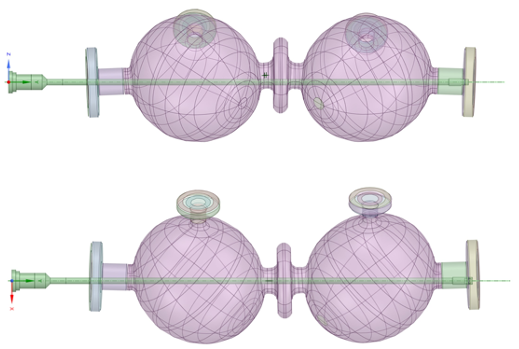}
    \caption{Simulation of the MAGO cavity installed on the HPR tool. In these two drawings the real cavity geometry is shown from two different angular perspectives, highlighting the necessity to straighten the cavity to reduce risk of interference between the cavity and the HPR wand.}
    \label{fig:hpr}
\end{figure}

A final crucial step before the RF cold test for an SRF cavity is often the high pressure rinsing (HPR). The HPR is performed to clean the inner surface of the cavity and remove contaminants such as dust or metal flakes that may have deposited on the cavity surface, and could negatively affect the cavity performance. A long and narrow HPR wand was used to avoid interference of the wand with the cavity walls while ensuring that the entire length of the cavity was washed; and a narrow spray head and small fan-type nozzles were used to minimize the transversal dimensions of the spray head and nozzles, out of concern for the narrow `beamtubes', see Fig.\,\ref{fig:hpr}.

\section{Cryogenic Tests}

In preparation for the first RF cold test, the cavity was assembled with three hook antennas to couple inductively to the magnetic field of the $\text{TE}_{011}$ mode. One input antenna with the intention to be critically coupled (with measured $\text{Q}_{ext1} = 1.6\times 10^{10}$) was installed on the `beamtube' axis, and the two field probes (with measured $\text{Q}_{ext 2,3} = 2-3\times 10^{12}$) were installed in the cell flanges on the transversal plane. Afterwards, the cavity underwent high pressure rinsing and evacuation. We will refer to the different antennas as `$1$' for the critically coupled antenna in cell 1, used as input coupler, `$2$' for the field probe in cell 1, `$3$' for the field probe in cell 2.
The first cryogenic test took place at Fermilab at 2\,K. The second and third cryogenic test took place at DESY at 4\,K and 2\,K. 
%The cavity was equipped with 8 temperature sensors, 5 flux gates, and Helmholtz coils for zero field cooldown in the Fermilab vertical insert. 
%-------------------------------------------------------------------------------
\subsection{RF Spectrum Measurements}

\subsubsection{Mode Splitting}
\label{sec:modesplitting}
%-------------------------------------------------------------------------------
As previously reported \cite{Fischer:2025}, it was necessary to plastically deform the two cells and tune their RF eigenfrequency differences from the order of $\mathcal{O}(\text{MHz})$ down to the intended $\mathcal{O}(\text{kHz})$ to make the two individual resonating cells behave like a coupled system, for the given cell-to-cell coupling $k_{cc}$. After tuning at room temperature, the two peaks merged as expected into a single broad resonance peak ($\mathcal{O}(\text{100~kHz})$), due to the low quality factor at room temperature, which causes the two resonances to overlap. Measuring the frequency of each cell individually, as uncoupled system, shows that the frequency difference at room temperature was between 4 and 7\,kHz. 

\begin{figure}[!ht]
     \includegraphics[width=\columnwidth]{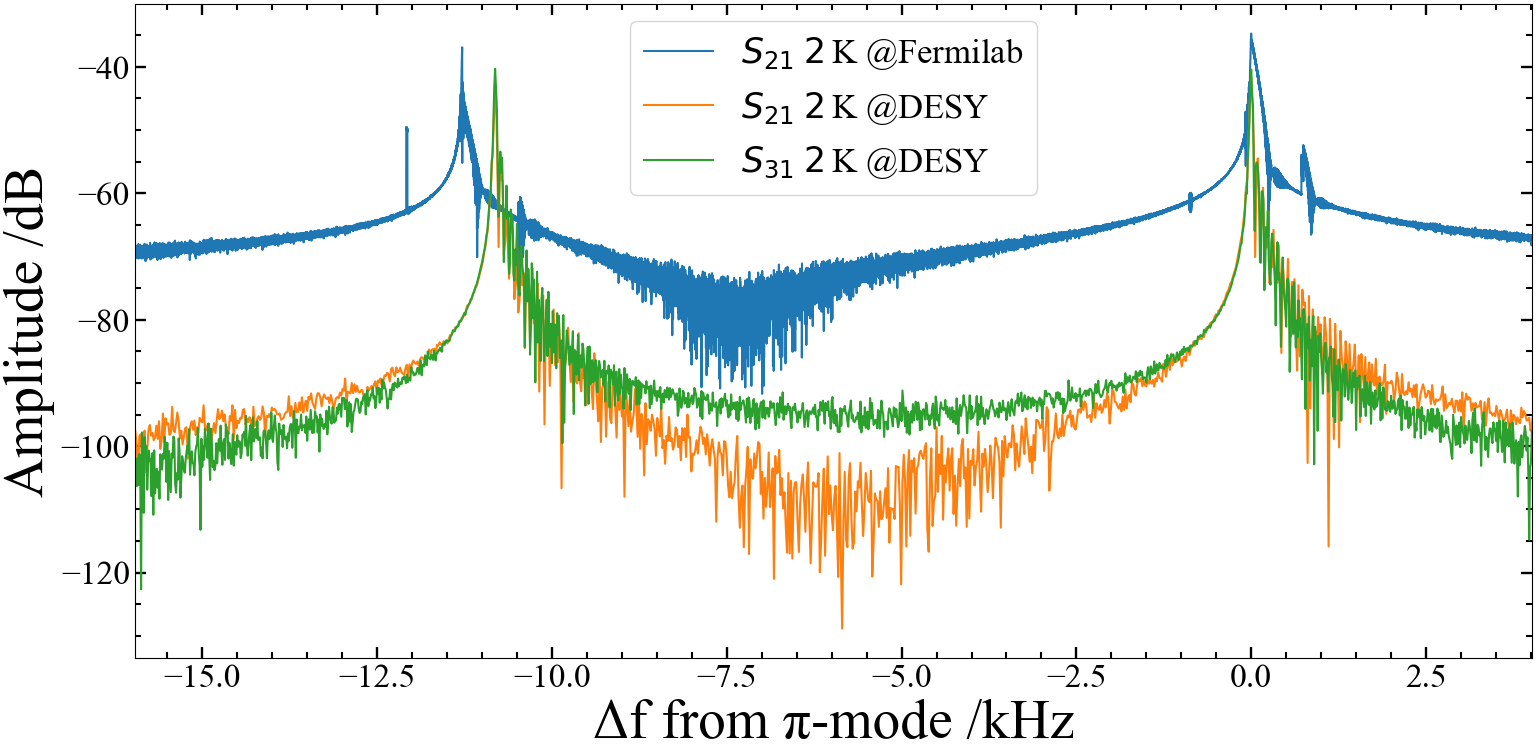}
     \caption{$S_{21}$ and $S_{31}$ transmission measurements around the $\text{TE}_{011}$ mode of the cavity at the two test facilities Fermilab and DESY. The frequencies are relative to the $\pi$-mode frequency, as differences in the absolute values of the eigenfrequencies are due to environmental changes. }
     \label{fig:modesplitting}
 \end{figure}

Because the two cells had been subjected to different mechanical stresses during tuning, and their fabrication histories were not fully known, it was difficult to predict how their eigenfrequencies would evolve during cooldown, which simultaneously will induce thermal contraction and a reduction in external pressure (from atmospheric to few mbar). This unpredictability raised the risk of the cells losing their symmetry, which would cause the frequency difference to deviate from the intended 10\,kHz. Fig.\,\ref{fig:modesplitting} presents the $S_{21}$ and $S_{31}$ transmission measurements of the cavity at 2\,K in the vicinity of the $\text{TE}_{011}$ mode, revealing a mode splitting of 11\,kHz. 

Furthermore, the configuration remained stable throughout transport from Fermilab to DESY. The slight discrepancy in mode splitting between the Fermilab and DESY spectra is attributed to minor environmental instabilities occurring between the multiple sweeps at Fermilab that were stitched together to span the full frequency range. The drift of the two eigenfrequencies during the frequency sweep also causes the smaller peaks in the Fermilab spectrum. To avoid these effects the DESY measurement was taken using a faster sweep with restricted frequency range and limited number of measurement points.

Between the two resonance peaks, the $S_{21}$ measurement exhibits a characteristic anti-resonance dip, which is a hallmark of coupled oscillator systems. It arises from destructive interference causing a sharp suppression of the transmitted signal at a frequency between the two eigenmodes. This anti-resonance can only be seen in the spectrum of the cell which hosts the input antenna, cell 1, and therefore is not seen in $S_{31}$, which is the spectrum taken from cell 2.

%-------------------------------------------------------------------------------
\subsubsection{\label{sec:phase_trans_func}Phase Transfer Function}

%-------------------------------------------------------------------------------

\begin{figure}[!htbp]

    \begin{subfigure}{\columnwidth}
    \centering
        \includegraphics[width=0.95\columnwidth]{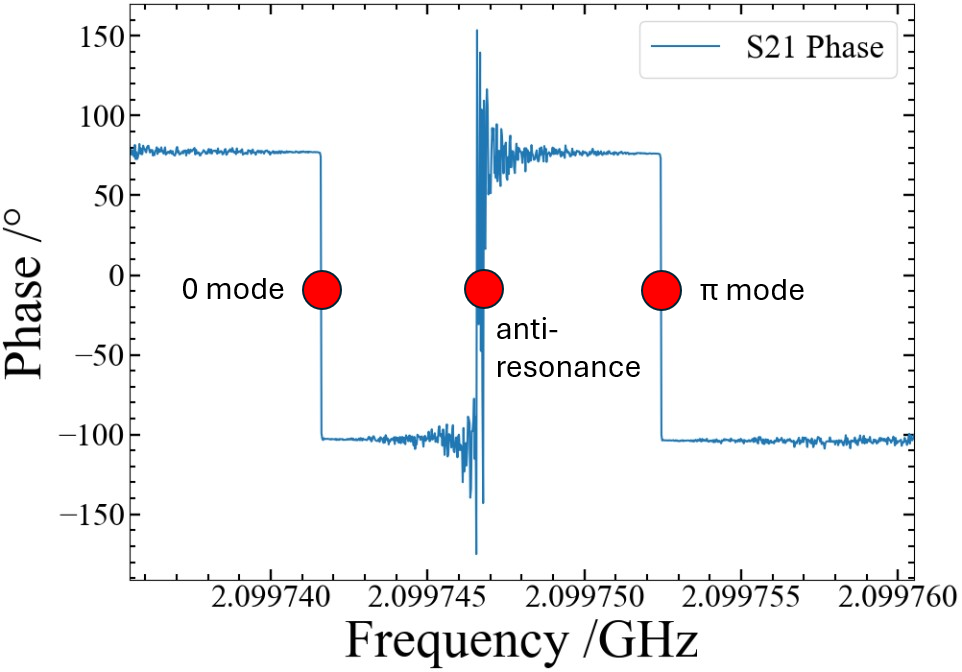}
        \caption{}
        \label{fig:phase_1}
    \end{subfigure}
    \hfill
    \begin{subfigure}{\columnwidth}
    \centering
        \includegraphics[width=0.95\columnwidth]{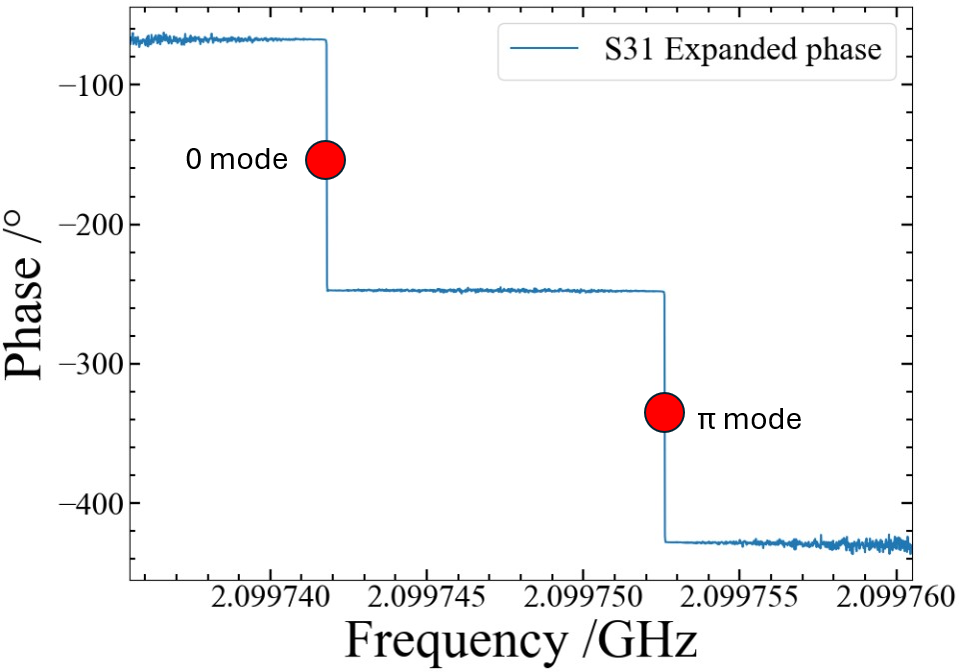}
        \caption{}
        \label{fig:phase_2}
    \end{subfigure}
    \label{fig:VNA}
    \caption{(a) VNA phase transfer function spectrum acquired from cell 1 and (b) cell 2 at 2\,K during the DESY cold test. The spectrum in (b) shows the expanded phase for simpler visualization of the "ladder-like" behaviour of the phase transfer function.}
\end{figure}

In Figs.\,\ref{fig:phase_1} and \ref{fig:phase_2}, the phase transfer functions measured at cell 1 and 2, respectively (corresponding to the $S_{21}$ and $S_{31}$ parameters), are presented. It can be seen that the signal measured from cell 1 shows a repetition of the same value three times: in the phase transition region at each eigenmode and at the `anti-resonance'. As we drive the cavity from cell 1, the `anti-resonance' is not present in cell 2. This results in a more conventional ladder-like transfer function. As we confirmed in the first test at Fermilab (later discussed in Section\,\ref{sec:QvE}), using the signal from cell 1 as reference for the low-level RF (LLRF) control system results in a system instability, since it is not possible for it to distinguish between the two modes. The frequency locking can be influenced by this, depending on the layout of the LLRF control system. 
 
\subsection{\label{sec:QvE}RF Quality Factor}

\begin{figure*}[!ht]
    \includegraphics[width=0.90\textwidth]{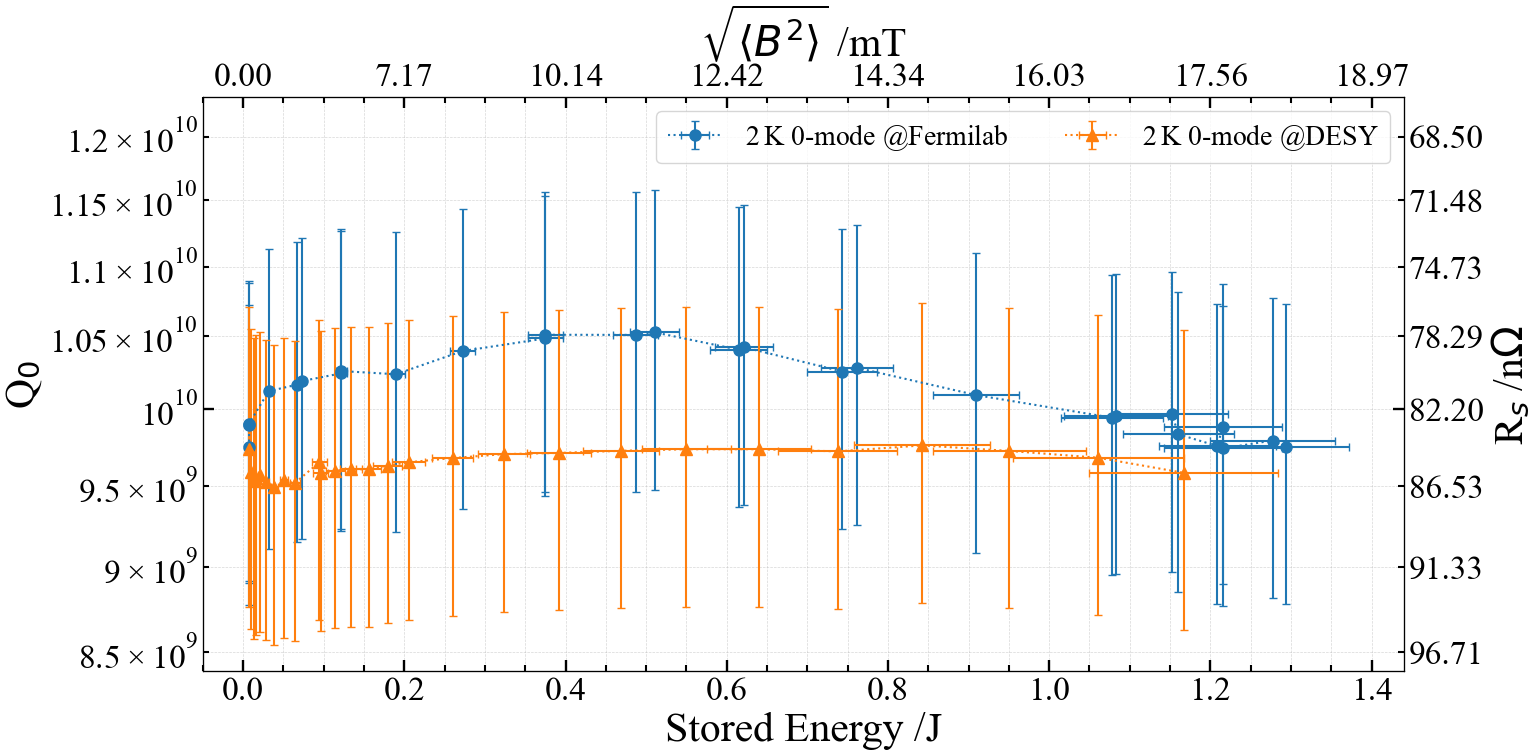}
    \caption{Intrinsic quality factor $Q_0$ as a function of stored energy for the $0$-mode, measured at Fermilab (blue circles) and DESY (orange triangles). On the top horizontal axis the corresponding average magnetic field in the cavity for a given stored energy is shown. On the right vertical axis, the conversion of $Q_0$ to surface resistance $Q_0=\frac{G}{R_s}$ is reported, where $G=823\,\Omega$ was obtained from eigenmode simulations. We assumed a conservative 10\% relative uncertainty \cite{powers2005theory, he2013uncertainty, melnychuk2014error, holzbauer2016systematic}}
    \label{fig:QvE}
\end{figure*}

The $0$-mode quality factor was measured at 2\,K at Fermilab and subsequently at 2\,K at DESY. Fig.\,\ref{fig:QvE} summarizes the performance of the mode measured from cell 1. The cavity exhibits high $Q_0$, in line with the performance of the previous MAGO prototype cavity\,\cite{Ballantini:2005am}. The $0$-mode quenched around 1.3\,J (hard quench). Measuring the antisymmetric $\pi$-mode proved challenging throughout both test campaigns. At Fermilab, measurements were restricted to low input power due to an instability in the phase-locked loop (PLL), caused by the phase transfer function of the system as mentioned in Section\,\ref{sec:phase_trans_func}. Even when exciting the the $\pi$-mode at low input power, a weak $0$-mode signal emerged above the noise floor, causing the PLL to preferentially lock onto it, rather than the $\pi$-mode, and making a clean $\pi$-mode measurement impossible. 

To address this issue during the second test at DESY, the reference signal was taken from cell 2 instead. However, an unexpected complication arose: as power was coupled into the $\pi$-mode, the $0$-mode became excited after a short delay, with oscillations in the $\pi$-mode amplitude closely mirrored by delayed oscillations in the $0$-mode amplitude. Unlike the Fermilab case, the control loop was not affected by phase ambiguity and therefore did not switch locking points, allowing both modes to remain simultaneously excited. Despite this improvement, the persistent excitation of the $0$-mode prevented a reliable measurement of the $\pi$-mode $Q_0$ at DESY as well.

These observations and previous experience with $9$-cell cavity testing pointed towards the presence of very low-energy field emission and/or multipacting that could transfer power from the $\pi$-mode to the $0$-mode. No radiation was detected by the radiation sensors outside the cryostat in either test. However, these sensors are positioned to optimally detect field emission from TESLA-type cavities operating in the accelerating mode, so the absence of a signal may simply reflect a geometrical limitation specific to the TE mode of this cavity geometry. Consequently, neither hypothesis can be confirmed nor excluded on this basis alone, though both remain compatible with the experimental observations and prior experience. Further simulations investigating the possibility of multipacting are presented in Section\,\ref{sec:multipact}.

In the course of these tests, additional noise sources were characterized, including RF input phase noise, amplifier noise, and microphonics; these results are not presented here, but will be discussed in a future dedicated publication.

As can be seen in Fig.\,\ref{fig:QvE} the results for the corresponding measurement in different facilities are compatible. The observed increase in the surface resistance measured at DESY can be explained by trapped flux which generates an additional contribution of surface losses, yet this possibility is still under investigation.
%-------------------------------------------------------------------------------
\subsection{\label{sec:dfdp}Pressure Sensitivity}

\begin{figure}
    \includegraphics[width=\columnwidth]{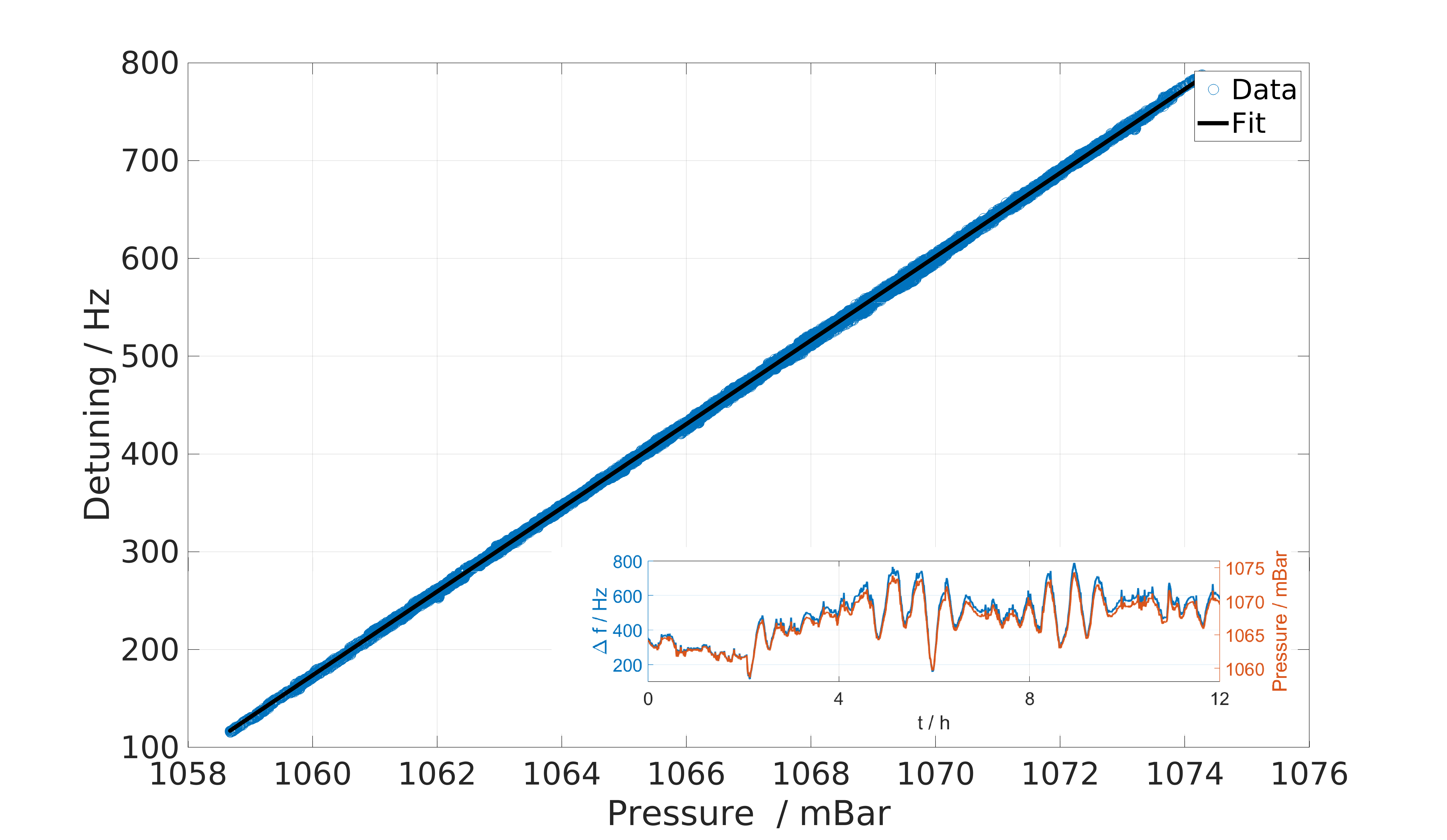}
    \caption{Frequency detuning $\Delta f$ vs. pressure as recorded during the 12\,h at 4\,K. A pressure sensitivity $df/dp$ of 42.8\,Hz/mbar is obtained. Inlet shows the respective timeseries of pressure (orange) and frequency detuning (blue).}
    \label{fig:dfdp}
\end{figure}

%-------------------------------------------------------------------------------
The high intrinsic quality factors achieved place stringent requirements on the stability of the resonance frequency, as even small environmental perturbations can detune the cavity beyond the bandwidth of the low-level RF control system. Therefore, during a 12\,h period of the 4\,K test at DESY, the frequency of the 0-mode was tracked by the RF driving system. During this time period, the pressure in the cryostat drifted by around 18\,mbar peak-to-peak. At the same time, the frequency of the 0-mode shifted up to 700\,Hz peak-to-peak as shown as inlet in Fig.\,\ref{fig:dfdp}. 
Plotting the frequency detuning vs. the pressure shows an almost perfect correlation, as shown in Fig.\,\ref{fig:dfdp}.
From this, the pressure sensitivity $df/dp$ of 42.8\,Hz/mbar for the cavity can be obtained. For comparison, the average  $df/dp$ of all 1.3\,GHz European XFEL cavities is $40\pm3$\,Hz/mbar and for the 3.9\,GHz European XFEL cavities $62\pm2$\,Hz/mbar. 
%-------------------------------------------------------------------------------
\subsection{\label{sec:mechmodes}Mechanical Quality Factor}

\begin{figure}[!htbp]
    \begin{subfigure}{\columnwidth}
      \centering
        \includegraphics[width=0.4\textwidth]{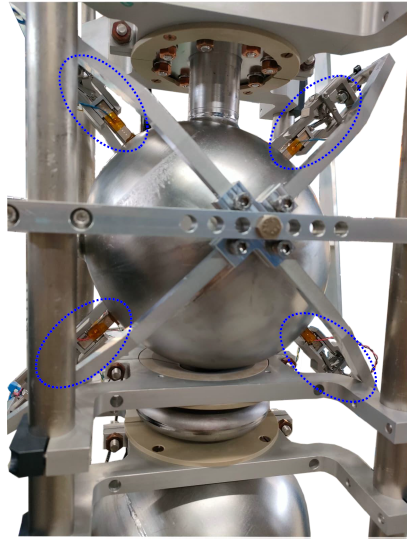}
        \caption{}
        \label{fig:Piezo}
    \end{subfigure}
    %\hspace{-2cm}  
    \begin{subfigure}{\columnwidth}
      \centering
        \includegraphics[width=\textwidth]{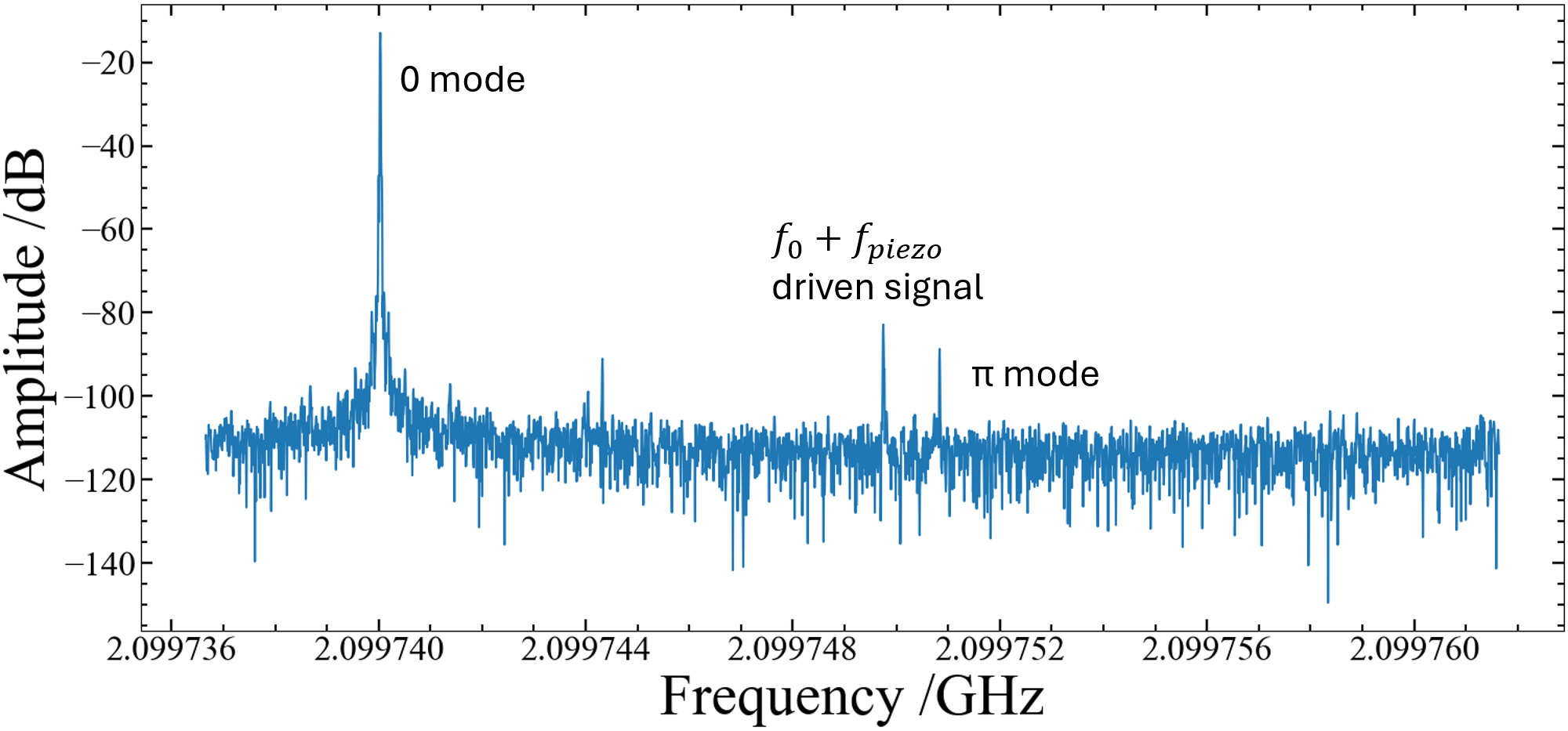}
        \caption{}
        \label{fig:piezo_peak}
      \label{fig:Piezo_and_spectrum}
    \end{subfigure}
    \caption{(a) One cell of the cavity was equipped with 4 piezo actuators (blue ellipses), where two opposite piezos were driven together in-phase and the other two then phase-shifted by 180° to generate a quadrupole vibration. (b) RF Spectrum taken from antenna 3 driving the cavity at the 0 mode and the piezoelectric actuators at $f_{piezo}$. A clear peak at the sum of the two frequencies appears in the spectrum.}
\end{figure}

%---
The mechanical eigenmodes and their respective mechanical quality factors $Q_m$ of the cavity at cryogenic temperatures are essential for predicting the expected signal and noise response. Until now, the values of $Q_m$ used in the literature for sensitivity predictions have not been based on measurements resembling realistic experimental setups, and as a result, tend to be overestimates. In contrast, our measurement more closely reflects the actual experimental conditions, as the cavity was not studied under idealized assumptions, but mounted in a frame and submerged in a helium bath. To conduct this study, we used the same piezo actuators used in the European XFEL for continuous cavity tuning, but deployed them directly in the liquid helium at 2\,K attached to the cavity with a custom structure. The piezos were tested, characterized, and calibrated at room temperature first \cite{Marconato2025}. The assembled cavity together with the four piezos is shown in Fig.\,\ref{fig:Piezo}.

\begin{figure}[!htbp]
	  \includegraphics[width=\columnwidth]{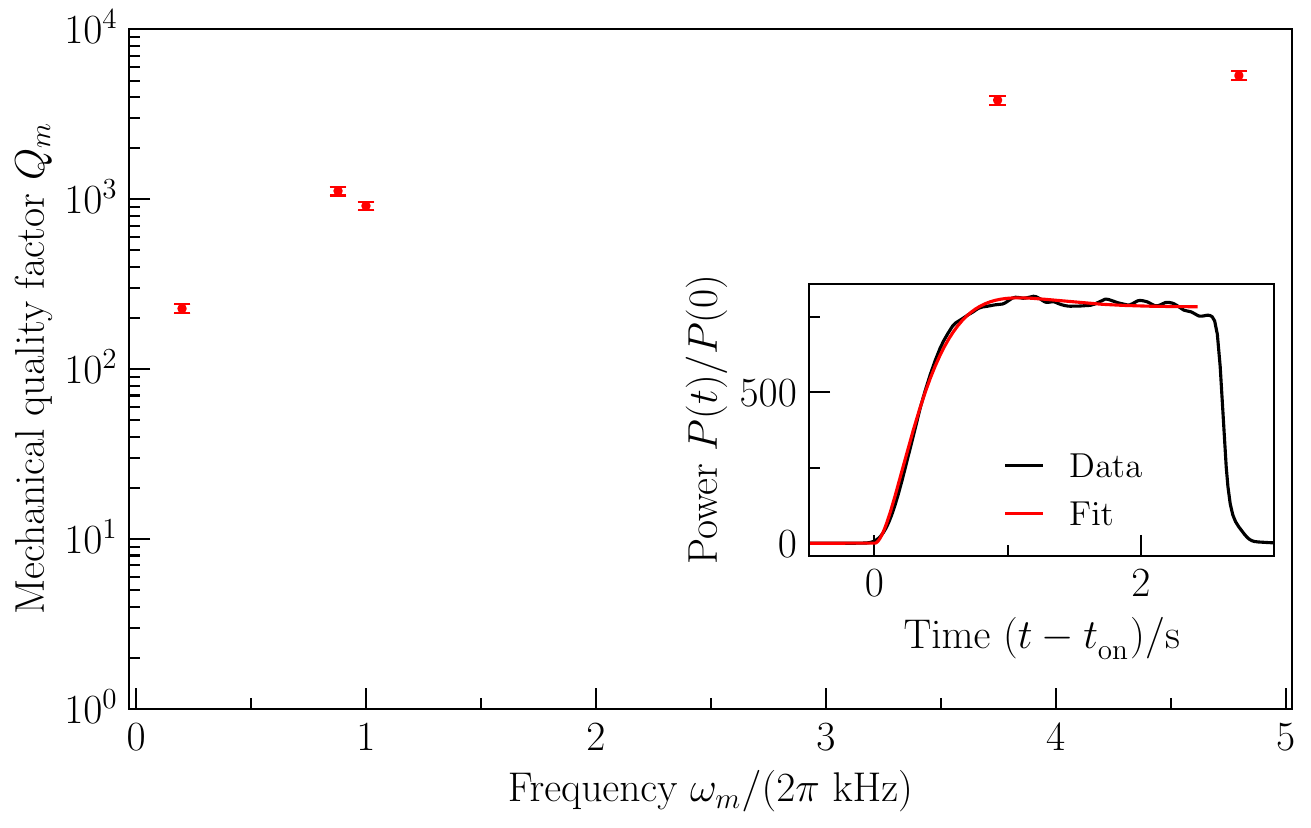}
	\caption{Mechanical quality factors obtained from fits to the normalized RF response of the cavity due to a mechanical pulse. Grey points show a fit assuming a perfectly resonant excitation and red points show an excitation with small detuning according to Eq. \eqref{eq:mechanical_loading_model}.}    	
	\label{fig:mecheigenmode}
\end{figure}

The piezos were excited with a fixed drive signal amplitude of 10\,V peak-to-peak and the frequency was swept from 0.1\,kHz to 10\,kHz while the cavity was driven at the $0$-mode. To identify possible mechanical resonances, the cavity RF spectrum was recorded during this phase; a representative example is shown in Fig.~\ref{fig:Piezo_and_spectrum}. The choice of mechanical eigenmode `candidates' was based on observations of significant increase of RF power in the piezo-driven RF peak with respect to neighboring frequencies. After positioning the piezo driving frequency on a mechanical eigenmode `candidate', a long pulse of about 10\,s was applied and the consequent RF power rise at the corresponding frequency ($\omega_0+\omega_d$) was measured. The inset in Fig.\,\ref{fig:mecheigenmode} shows the RF response in the time domain for such an excitation at 3.7\,kHz. 
In order to obtain the quality factor, we fit to the RF response, assuming a single harmonic oscillator with frequency $\omega_m$, excited with a monochromatic drive at frequency $\omega_d$
\begin{equation}\label{eq:mechanical_loading_model}
    P(t)=P_0\left|1-e^{i(\omega_m-\omega_d)t}e^{-\frac12\frac{\omega_m}{Q_m}t}\right|^2\,,
\end{equation}
where the drive is taken to be switched on at time $t=t_\text{on}=0$. A derivation of this formula and a discussion of the approximation that a single harmonic oscillator is being excited in the presence of nearby EM resonances can be found in the Appendix\,\ref{app:harmonic_oscillators}. The resulting quality factors are shown in Fig.\,\ref{fig:mecheigenmode} and are obtained by manually setting the switch-on time $t_\text{on}$, using the known drive frequency $\omega_d$ and fitting to $P_0$, $\omega_m$ and $Q_m$. To record the time domain traces a spectrum analyzer in `zero-span mode' was used. In this mode, the instrument monitors a fixed frequency of the RF spectrum instead of sweeping over multiple frequencies, and while the piezo driving frequency was stable, the cavity 0-mode frequency was affected by small fluctuations, as expected from small pressure variations and microphonics. Therefore it is plausible that during the recording of a mechanical mode decay, the piezo-driven RF peak was slightly drifting from its frequency $\omega_0+\omega_d$. Hence, the fit quality improves when a possible detuning $\omega_d-\omega_m$ is taken into account. The relative statistical errors on $Q_m$ from the fit range from $0.3\%-0.4\%$. However, a larger systematic error arises due to manually choosing the switch-on time. By performing the fits for different acceptable times, we estimate a $\sim6\%$ relative error for the fit. Another systematic error emerges due to the drifing of the RF resonances during the $\mathcal{O}(s)$ long measurement, which affects the RF power level of the mechanical excitation. However, since the mechanical modes of interest are on the tail of the pump resonance, the fluctuations are expected to be negligible compared to the other errors. 

The observed values of $Q_m\sim 10^3$ are in line with values obtained in comparable setups \cite{Ballantini2003}, however far below the value $Q_m\sim 10^6$ from idealized theoretical values for bare niobium cavities in liquid helium \cite{Ballantini:2005am}, which have been assumed in previous works \cite{Berlin:2023grv, Fischer:2025}. 
%-------------------------------------------------------------------------------
\section{\label{sec:multipact}Numerical Multipacting Study}
%-------------------------------------------------------------------------------

It is known that multipacting in a multicell cavity system can cause the excitation of unwanted modes \cite{volkov_monopole_2010, Kreps2009}, similarly to what was observed in the tests previously described in Section~\ref{sec:QvE}. To investigate this effect, multipacting simulations replicating the experimental conditions were conducted.

Multipacting is a well-known and extensively studied phenomenon for several cavity geometries, as it poses a critical limitation in particle accelerators. In contrast, multipacting in geometries other than accelerating structures is less understood, and its onset is difficult to predict without simulations. Multipacting can generally be divided into two types: one-point and two-point. The second type is the dominant mechanism in TESLA-type cavities, while the first one is usually not a concern as most common modern cavity geometries mitigate them by design choices. This is possible as multipacting is a resonant phenomenon that requires specific stability conditions to be satisfied. These conditions in case of two-point multipacting give rise to `energy barriers': some specific EM field amplitudes at which the resonant conditions are met. In the measurements conducted on our cavity, no such barriers could be identified, and the stored energy required to trigger multipacting (assuming it to be the cause of the observed behaviour) was extremely low. These considerations point towards the presence of one-point multipacting.

\begin{figure}[!ht]
    \begin{subfigure}{\columnwidth}
    \centering
         \includegraphics[width=0.9\textwidth]{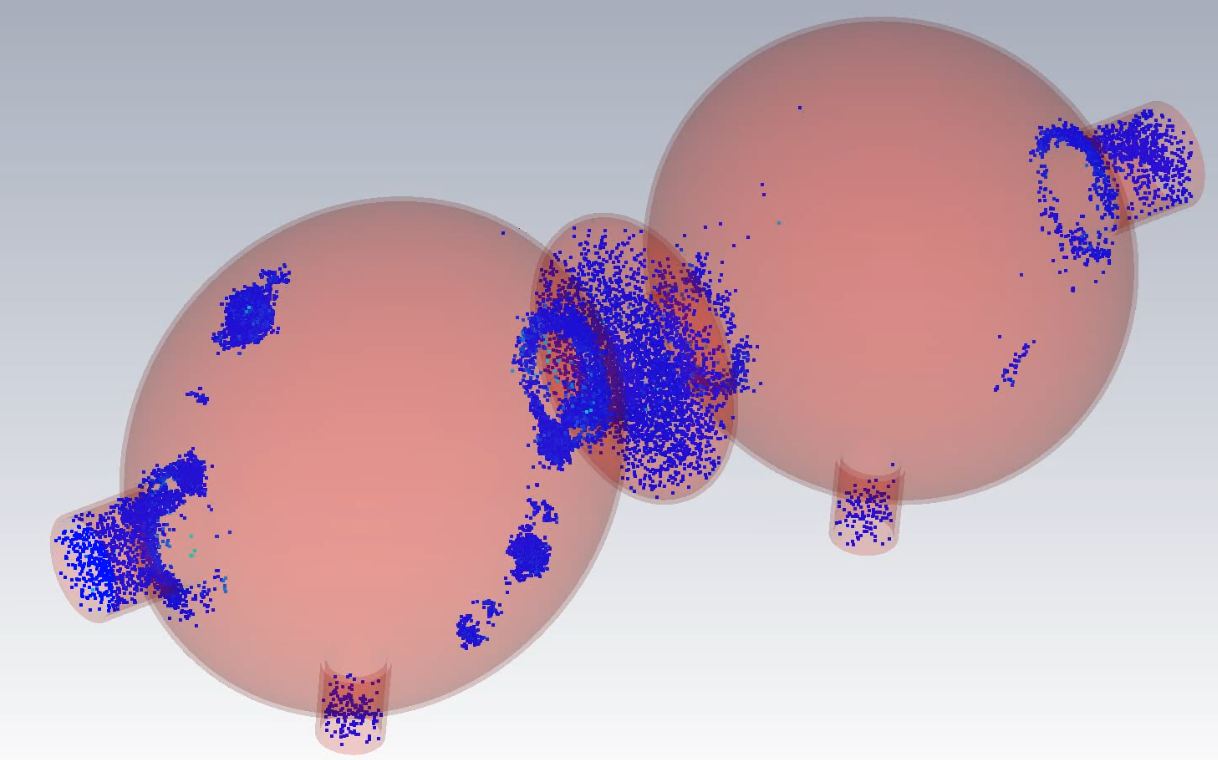}
         \caption{}
         \label{fig:multipact_screenshot}
    \end{subfigure}

    \bigskip

    \begin{subfigure}{\columnwidth}
   
        \includegraphics[width=1\textwidth]{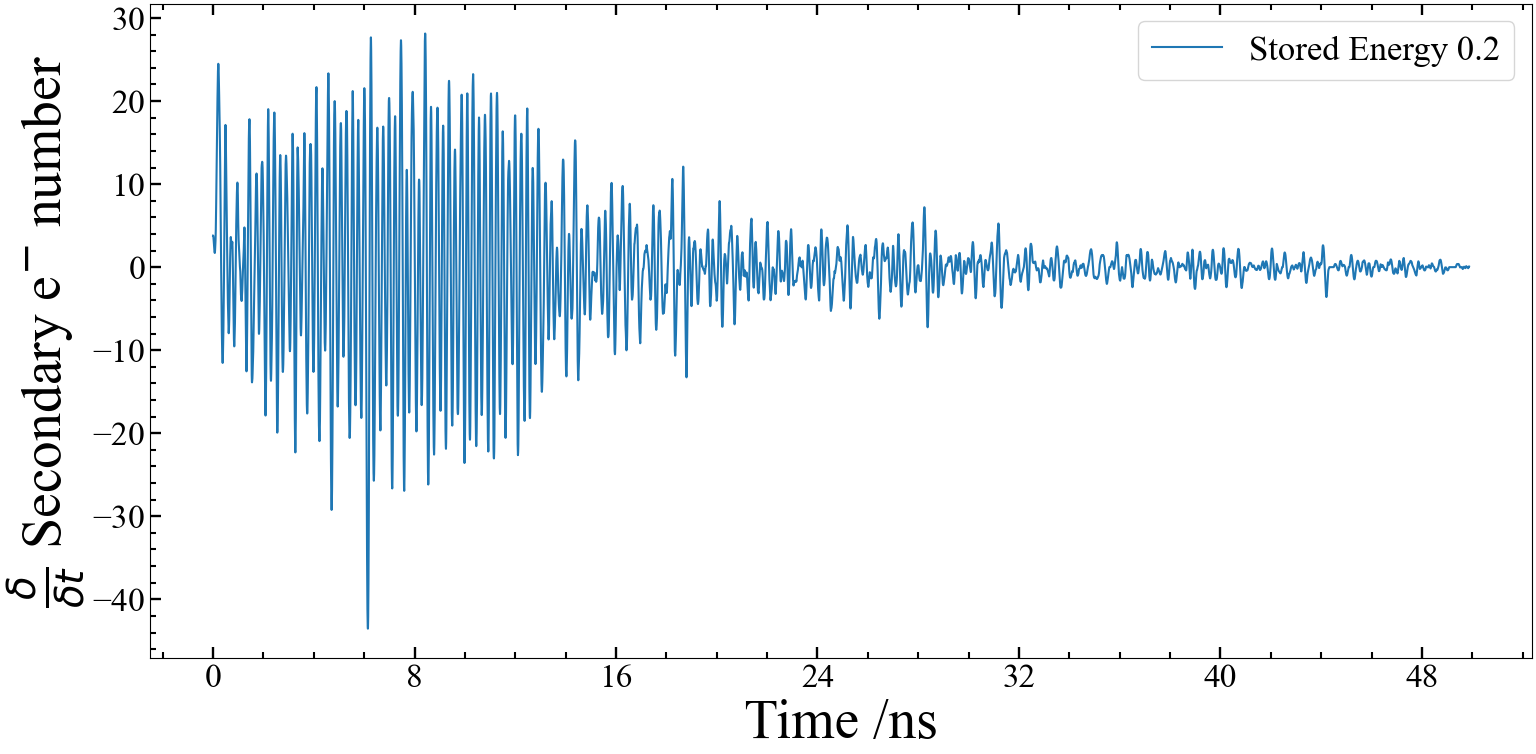}
        \caption{}
        \label{fig:secondary_e_low_E}
    \end{subfigure}
    \hfill
    \begin{subfigure}{\columnwidth}
  
        \includegraphics[width=1\textwidth]{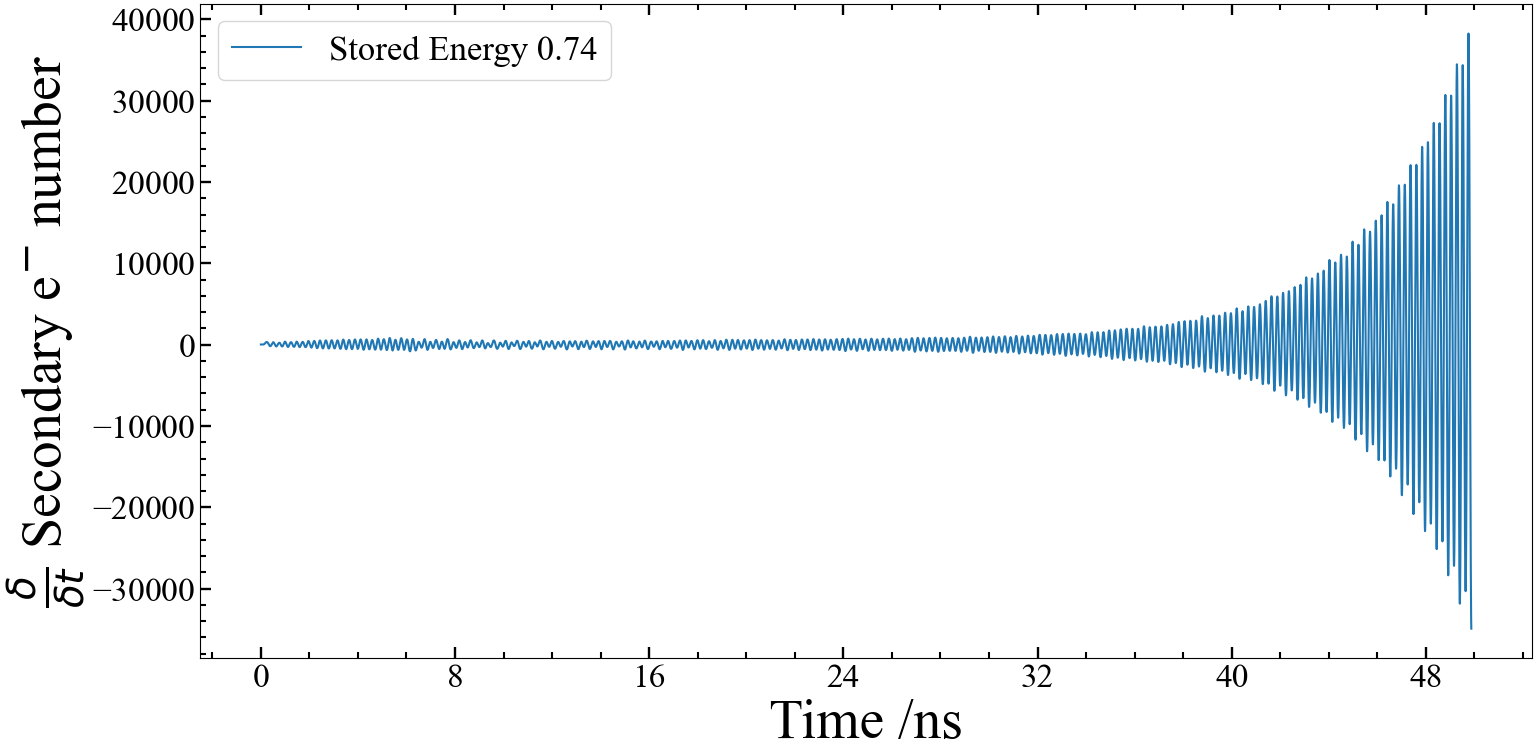}
        \caption{}
        \label{fig:secondary_e_high_E}
    \end{subfigure}
    \label{fig:Multipacting}
    \caption{(a) Example of multipacting areas on the inner cavity surface for stored energy of 0.74\,J. (b)(c)Time derivative of the number of secondary electrons emitted from the cavity inner surface calculated from time-domain simulations of multipacting with (b) 0.2\,J and (c) 0.74\,J stored energy. With lower stored energy the growth decreases drastically in a short amount of time, while with higher energy it can increase exponentially even after many RF cycles.}
\end{figure}

A result of the simulations for our cavity is shown in Fig.\,\ref{fig:multipact_screenshot}. The 3D model shows the points of impact of initially randomly distributed electrons on the inner surface of the cavity when the system is driven at its RF eigenmode with a stored energy of 0.74\,J. The software can automatically detect the onset of multipacting by fitting an exponential to the time derivative of the number of secondary electrons produced, and stops the simulation if the fit is successful. Multipacting was always detected for stored energy values of 0.8\,J or higher, while it was never observed for stored energies of 0.6\,J or lower. Fig.\,\ref{fig:secondary_e_low_E} and \ref{fig:secondary_e_high_E} show the time derivative of the number of electrons produced for 0.2\,J and 0.74\,J respectively. As it can be seen for the lower stored energy, the electron growth decreases quickly with time. For stored energies between 0.6\,J and 0.8\,J multipacting does not trigger within a few RF cycles (low order multipacting $\sim1$), but still produces an exponential secondary electron emission at a later time. This energy interval is a transition regime in which higher-order multipacting occurs. Higher-order multipacting is usually considered less dangerous for accelerator operation, and usually it is not the cause of early quench of cavities, but for sensing applications could still be an unwanted source of noise to be eliminated.

The simulations are in agreement with the observed behaviour, as no well defined multipacting energy barriers have been identified. On the contrary, multipacting appears to be always present above a certain starting energy threshold. The starting threshold of 0.6\,J obtained in the simulations is not confirmed experimentally, however, this can be explained as the multipacting simulations were conducted on the ideal cavity geometry with uniform and smooth inner surface, without taking into account the cavity imperfections, defects and surface roughness. Inner surface defects can indeed lower the multipacting activation threshold by increasing the secondary emission yield (SEY) compared to smooth defectless niobium. The presence of multipacting above a certain threshold, together with the trajectories extracted from time domain simulations, confirms that at each location showed in Fig.\,\ref{fig:multipact_screenshot} is indeed a one-point multipacting.

Although the results do not provide conclusive proof of multipacting during the Fermilab and DESY tests, they identify it as a credible possibility that should be examined through dedicated follow-up measurements. Confirming this effect would have important consequences for the design of future cavity geometries. 

%-------------------------------------------------------------------------------
\section{\label{sec:conclusion}Conclusion}

%-------------------------------------------------------------------------------
The cryogenic RF characterization of the cavity demonstrated the successful transfer of established SRF cavity preparation and testing methods to a novel cavity geometry with significant mechanical and electromagnetic uncertainties. Due to the unknown fabrication and treatment history of the cavity, together with the presence of thin-wall regions, the surface preparation strategy followed a deliberately cautious approach. A light flash BCP, moderate heat treatment, and adapted HPR procedure were successfully implemented while minimizing risks associated with the unconventional geometry.

The room-temperature mechanical tuning campaign proved successful, as measurements of the individual eigenfrequencies of the cells indicated a frequency difference of approximately 5\,kHz. The coupled-cell system reached a mode splitting of approximately 11\,kHz at 2\,K, close to the intended operating regime. Importantly, no measurable change in the mode splitting was observed after the transportation of the cavity between Fermilab and DESY, demonstrating good mechanical stability of the tuned configuration.
The RF spectrum measurements further revealed important features for the operation of the LLRF control system. When the pickup signal was taken from cell~1, the phase transfer function exhibited repeated phase values for the two eigenmodes and the anti-resonance, leading to ambiguities in the phase-locked loop operation. In contrast, signals acquired from cell~2 showed a unique phase evolution for each mode, enabling stable frequency locking and providing important operational insights for future control system implementations.

Measurements of the intrinsic quality factor showed good agreement between independent cryogenic tests performed at Fermilab and DESY, indicating reproducible cavity performance and confirming the effectiveness of the applied surface chemistry. The achieved performance is also consistent with previous MAGO prototype cavities. Nevertheless, persistent difficulties in stably operating and characterizing the $\pi$-mode were observed. Experimental observations suggest that low-energy multipacting may be responsible for the unwanted coupling between modes and the instability of the locking system. Initial multipacting simulations support this interpretation and indicate that one-point multipacting could occur in this cavity geometry over a broad energy range. 

It should be noted that two other observations can also be explained by multipacting. The curve measured at Fermilab, shown in Fig.\,\ref{fig:QvE}, starts to show an increase in the RF losses just around 0.6\,J, coinciding with the numerical threshold at which multipacting should start. The curve measured at DESY starts at a higher surface resistance, and, hence, may be less sensitive to additional losses caused by multipacting. The second observation is the similar quench field in the cavity tested here, at 1.2\,J, and in the previous prototype cavity test, at 1.6\,J, with the test results reported in \cite{Ballantini:2003nt}. There, an RF breakdown in the antenna was reported as the most likely issue and that can fit with multipacting, but no study on that geometry was conducted. Given that the region where multipacting was identified in the simulations presented in this study is geometrically identical between the two cavities, it is possible that the previous cavity was also limited by multipacting, resulting in a similar breakdown field. A dedicated measurement will be conducted to fully clarify the origin of these effects and either confirm or rule out multipacting. This is a significant insight, as it has consequences for the geometry of the next detector cavity.

The LLRF system demonstrated good frequency tracking behavior of the narrow resonance during long-term operation. A pressure sensitivity of $df/dp = 42.8\,\mathrm{Hz/mbar}$ was measured, comparable to values observed in conventional SRF cavities, proving the feasibility of our concepts for future LLRF-based detector operation and stabilization.

Finally, a first cryogenic characterization of the mechanical eigenmodes was successfully performed using immersed piezo actuators at 2\,K. Mechanical quality factors were extracted from RF measurements and found to be of the order of $Q_m \sim 10^3$, substantially lower than the idealized values commonly assumed in the literature. 

As mentioned, the measured mechanical quality factors characterize the entire system, including both the cavity and its supporting frame, particularly for low-frequency modes. Preliminary simulations indicate that the influence of the frame decreases at higher frequencies, which may explain the observed increase of $Q_m$ with frequency, assuming that the bare cavity exhibits intrinsically higher quality factors. This additional damping introduced by support structures or frames is often neglected in sensitivity estimates, although it can significantly affect the achievable experimental reach. In the case that external vibrations dominate the noise budget, as we expect for a first GW search with the MAGO prototype, the GW sensitivity does not depend on $Q_m$. However, in a future setup, large quality factors can be necessary to increase the strain sensitivity $h_\text{min}\propto Q_m^{-1}$ when EM noise sources dominate until the noise floor of thermal vibrations is reached, where $h_\text{min}\propto Q_m^{-1/2}$. These results provide an important experimental benchmark for future sensitivity estimates and emphasize the relevance of mechanical losses and suspension conditions in realistic detector implementations.

% If in two-column mode, this environment will change to single-column
% format so that long equations can be displayed. Use
%\begin{widetext}
%\end{widetext}

\appendix*
\section{\label{app:harmonic_oscillators}RF Signal of Cavity Vibration}
The aim of this section is to describe how a vibration of the cavity walls excited by the piezos causes an electromagnetic signal while the 0-mode is excited. This allows us to demonstrate that the decay measurement in Section\,\ref{sec:mechmodes} is a clean probe of the mechanical quality factor even though nearby electromagnetic resonances complicate the signal transfer function. We will proceed by deriving Eq. \eqref{eq:mechanical_loading_model} and clarifying under which assumptions it is an accurate description.

The cavity is excited with a magnetic field $\bar{\bm{B}}(t,\bm{x})=B_0\bm{B}_0(\bm{x})\cos{\omega_0 t}$, where $B_0$ is the root mean square volume average and $\bm{B}_0(\bm{x})$ is the field distribution of the 0-mode, normalized so that $\int dV|\bm{B}_0(\bm{x})|^2=V$. The perturbation of the pump mode due to the piezo's force can then be written as $\delta\bm{B}(t,\bm{x})=b_0(t)\bm{B}_0(\bm{x})$. For simplicity, we will only consider the signal arising in the 0-mode, since it is closer in frequency to the mechanical modes in Fig.\,\ref{fig:mecheigenmode} and therefore dominant. However, the analysis can easily be extended to include more EM modes. 

Cavity perturbation theory then yields the result \cite{Gue:2026kga}
\begin{equation}\label{eq:b_0_piezo_solution}
    b_0(\omega_0+\omega)=\frac{C_mB_0}{2V^{1/3}M}\frac{\omega_0^2\,f^\text{piezo}_m(\omega)}{R_m(\omega)R_0(\omega+\omega_0)-\frac{C_m^2\omega_0^2U_0}{2MV^{2/3}}}\,,
\end{equation}
with the resonance functions
\begin{equation}
    R_{0/m}(\omega)=\omega_{0/m}^2-\omega^2+i\omega \gamma_{0/m},
\end{equation}
the damping coefficients $\gamma_{0/m}=\omega_{0/m}/Q_{0/m}$, the mass of the cavity walls $M$, the stored energy in the pump mode $U_0$, and the coupling coefficient between pump mode and a mechanical eigenmode $\bm{U}_m(\bm{x})$
\begin{equation}
    C_m=\frac{1}{V^{2/3}}\int_{\partial V} d\bm{A}\cdot\bm{U}_m(|\bm{B}_0|^2-|\bm{E}_0|^2)\,.
\end{equation}
The mechanical eigenmodes are normalized so that $\int_{V_w} dV\,|\bm{U}_m|^2=V_w$ where $V_w$ is the volume of the cavity walls. The strength $f_m^\text{piezo}$ with which the piezos excite a mechanical mode is given by
\begin{equation}\label{eq:piezo_force}
    f_m^\text{piezo}(t)=\sum_pP_p(t)\int_{A_p}d\bm{A}\cdot\bm{U}_m\,,
\end{equation}
where $p$ indexes the four piezos with the surface area $A_p$ they act on, and the pressure $P_p(t)$ they exert along the cavity's surface normal. For $f_m^\text{piezo}(t<0)=0$ and $f_m^\text{piezo}(t\geq0)=F_me^{i\omega_d t}$, the Fourier transform of Eq. \eqref{eq:piezo_force} is given by
\begin{equation}
f_m^\text{piezo}(\omega)=\lim_{\epsilon\to0}\frac{-iF_m}{\omega-\omega_d-i\epsilon}\,,
\end{equation}
where we have moved the pole at $\omega=\omega_d$ into the positive complex plane to simplify our integration later.

The term $\propto U_0$ in the denominator of Eq. \eqref{eq:b_0_piezo_solution} arises due to back-action of the pump field on the wall vibration. However, since we generically expect $C_m\lesssim0.1$ for mechanical modes in this frequency regime \cite{Fischer:2025}, we have
\begin{equation}
    \frac{C_m^2\omega_0U_0}{2MV^{2/3}\omega_m^2}\ll\gamma_{m,0}\,,
\end{equation}
which allows us to neglect the back-action term.\footnote{This approximation is not expected to hold anymore for higher couplings $C_m\sim1$ and larger stored energies $U_0$ than in this test setup. However, the following analysis would still remain valid, as long as the resonance frequencies and damping coefficients of the coupled EM-mechanical system are used instead.}

To find the evolution of $b_0$ in the time domain, we can calculate the inverse Fourier transform of Eq. \eqref{eq:b_0_piezo_solution} 
\begin{align}\label{eq:b_t_inv_Fourier}
    b_0(t)&=\frac{C_m\omega_0F_m}{4MV^{1/3}}B_0e^{i\omega_0t} \\
    &\cdot\int\frac{d\omega}{2\pi i}\frac{e^{i\omega t}}{(\omega-\omega_m^+)(\omega-\omega_m^-)(\omega-i\frac{\gamma_0}{2})(\omega-\omega_d-i\epsilon)}\,,\nonumber
\end{align}
where we have defined the pole frequencies $\omega_m^{\pm}=\pm\omega_m+i\frac{\gamma_m}{2}$ and used $\omega\ll\omega_0$, as well as $\omega_m\gg\gamma_{m,0}$. The integral in Eq. \eqref{eq:b_t_inv_Fourier} makes it evident that the time evolution of $b_0(t)$ for $\omega_d\approx\omega_m$ is not only determined by the mechanical resonance, but the entire transfer function $b_0(\omega)/f_m^\text{piezo}(\omega)$ is being probed. However, frequency components away from the driving frequency are suppressed $\propto(\omega-\omega_d)^{-1}$. 

After solving the integral in Eq. \eqref{eq:b_t_inv_Fourier} by using the residue theorem, we find
\begin{widetext}
\begin{align}
    b_0(t)=&\,\frac{C_m\omega_0F_m}{4\omega_dMV^{1/3}}\frac{B_0e^{i\omega_0t}}{\omega_d^2-\omega_m^2-i(\omega_d+\omega_m)\frac{\gamma_m}{2}}\Bigg[e^{i\omega_dt}-\frac{\omega_d(\omega_d+\omega_m)}{2\omega_m^2}e^{i\omega_mt-\frac{\gamma_m}{2}t}\nonumber\\
    &+\frac{\omega_d-\omega_m-i\frac{\gamma_m}{2}}{\omega_m}\left(\frac{\omega_d+\omega_m}{\omega_m}e^{-\frac{\gamma_0}{2}t}-\frac{\omega_d}{2\omega_m}e^{-i\omega_mt-\frac{\gamma_m}{2}t}\right)\Bigg]\,,
\end{align}
\end{widetext}
where we have used again that $\omega_m\gg\gamma_0$ and $\omega_d\gg\gamma_0$. The term $\propto e^{-\frac{\gamma_0}{2}t}$ arises due to the 0-mode resonance, however is suppressed by a factor $1/Q_m$ against the mechanical decay $\propto e^{-\frac{\gamma_m}{2}t}$ under the condition that $|\omega_d-\omega_m|\lesssim\gamma_m$. Since the piezo drive was tuned to be near mechanical resonances, this assumption is well justified, and any contributions from the EM resonance can be neglected. The term $\propto e^{-i\omega_mt}$ can be neglected on the same grounds, so that we can write the solution normalized by its steady state value as
\begin{equation}
    \frac{b_0(t)}{b_0(t\to\infty)}\approx 1-\frac{\omega_d(\omega_m+\omega_d)}{2\omega_m^2}e^{i(\omega_m-\omega_d)t-\frac{\gamma_m}{2}t}\,.
\end{equation}
Thus, after using $|\omega_m-\omega_d|\ll\omega_m$ and taking the absolute value, we arrive at Eq. \eqref{eq:mechanical_loading_model} used in the main text to find the mechanical quality factor.

\begin{acknowledgments}
For the loan of the {\tt PACO-2GHz-variable} cavity, we thank the Istituto Nazionale di Fisica Nucleare, Italy.\\
The authors thank Asher Berlin, Thorsten Buettner, Sergio Calatroni, Sebastian Ellis, Gianluca Gemme, Roni Harnik, Denis Kostin, Frank Ludwig, Cornelius Martens, Andreas Ringwald, Holger Schlarb, Udai Raj Singh, Hans Weise and Mateusz Wiencek for their support and useful discussions.
This material is based upon work supported by the U.S. Department of Energy, Office of Science, National Quantum Information Science Research Centers, Superconducting Quantum Materials and Systems Center (SQMS), under Contract No. 89243024CSC000002. Fermilab is operated by Fermi Forward Discovery Group, LLC under Contract No. 89243024CSC000002 with the U.S. Department of Energy, Office of Science, Office of High Energy Physics. The project is also funded/acknowledges support by the Deutsche Forschungsgemeinschaft (DFG, German Research Foundation) under Germanys Excellence Strategy - EXC 2121 `Quantum Universe' - 390833306.
\end{acknowledgments}

\bigskip
C.D., B.G., T.K., G.M., K.P. and M.W. contributed equally to this work.

\bibliography{biblio}

\end{document}